\documentclass[review]{elsarticle}

\usepackage{lineno,hyperref}
\modulolinenumbers[5]

\journal{Physics Letters B}

\usepackage{graphics} 
\usepackage{graphicx} 
\usepackage{epstopdf}
\usepackage{color}
\usepackage{scrtime}
\usepackage{amssymb}
\usepackage{amsmath}
\usepackage{amsfonts}
\usepackage{cleveref}
\usepackage{multirow}
\usepackage{pdflscape}
\usepackage[ulem=normalem]{changes}

\begin{document}

\begin{frontmatter}

\title{Electromagnetic properties of $^{21}$O for benchmarking nuclear Hamiltonians}


\author[IKP]{S.~Heil}
\author[York,IKP]{M.~Petri\corref{mycorrespondingauthor}}
\cortext[mycorrespondingauthor]{Corresponding author}
\ead{marina.petri@york.ac.uk}
\author[IKP]{K.~Vobig}
\author[NSCL,MSU]{D.~Bazin}
\author[NSCL]{J.~Belarge\fnref{belarge}}
\fntext[belarge]{J. Belarge is currently an MIT Lincoln Laboratory employee. No Laboratory funding or resources were used to produce the result/findings reported in this publication.}
\author[NSCL,Lowell]{P.~Bender}
\author[NSCL,MSU]{B.~A.~Brown}
\author[NSCL,MSU]{R.~Elder}
\author[NSCL,MSU]{B.~Elman}
\author[NSCL,MSU]{A.~Gade}
\author[York]{T.~Haylett}
\author[TRIUMF]{J.~D.~Holt}
\author[IKP]{T.~H\"uther}
\author[IKP]{A.~Hufnagel}
\author[NSCL,MSU]{H.~Iwasaki}
\author[NSCL]{N.~Kobayashi}
\author[NSCL,MSU]{C.~Loelius}
\author[NSCL,MSU]{B.~Longfellow}
\author[NSCL,MSU]{E.~Lunderberg} 
\author[IKP]{M.~Mathy}
\author[Tokyo]{J.~Men\'endez}
\author[York]{S.~Paschalis}
\author[IKP]{R.~Roth}
\author[IKP,EMMI,Max]{A.~Schwenk}
\author[IKP]{J.~Simonis}
\author[IKP]{I.~Syndikus}
\author[NSCL]{D.~Weisshaar}
\author[NSCL,MSU]{K.~Whitmore}

\address[IKP]{Institut f\"ur Kernphysik, Technische Universit\"at Darmstadt, 64289 Darmstadt, Germany}
\address[York]{Department of Physics, University of York, Heslington, York, YO10 5DD, UK}
\address[NSCL]{National Superconducting Cyclotron Laboratory, Michigan State University, East Lansing, Michigan 48824, USA}
\address[MSU]{Department of Physics and Astronomy, Michigan State University, East Lansing, Michigan 48824, USA}
\address[Lowell]{Department of Physics, University of Massachusetts Lowell, Lowell, Massachusetts 01854, USA}
\address[TRIUMF]{TRIUMF 4004 Wesbrook Mall, Vancouver, British Columbia V6T 2A3, Canada}
\address[Tokyo]{Center for Nuclear Study, The University of Tokyo, Tokyo, Japan}
\address[EMMI]{ExtreMe Matter Institute EMMI, GSI Helmholtzzentrum f\"ur Schwerionenforschung GmbH, 64291 Darmstadt, Germany}
\address[Max]{Max-Planck-Institut f\"ur Kernphysik, 69117 Heidelberg, Germany}

\begin{abstract}

The structure of exotic nuclei provides valuable tests for state-of-the-art nuclear theory. 
In particular electromagnetic transition rates are more sensitive to aspects of nuclear forces and many-body physics than excitation energies alone.
We report the first lifetime measurement of excited states in $^{21}$O, finding $\tau_{1/2^+}=420^{+35}_{-32}\text{(stat)}^{+34}_{-12}\text{(sys)}$\,ps. 
This result together with the deduced level scheme and branching ratio of several $\gamma$-ray decays are compared to both phenomenological shell-model and ab initio calculations based on two- and three-nucleon forces derived from chiral effective field theory.
We find that the electric quadrupole reduced transition probability of $\rm B(E2;1/2^+ \rightarrow 5/2^+_{g.s.}) = 0.71^{+0.07\ +0.02}_{-0.06\ -0.06}$~e$^2$fm$^4$, derived from the lifetime of the $1/2^+$ state, is smaller than the phenomenological result where standard effective charges are employed, suggesting the need for modifications of the latter in neutron-rich oxygen isotopes. 
We compare this result to both large-space and valence-space ab initio calculations, and by using multiple input interactions we explore the sensitivity of this observable to underlying details of nuclear forces.

\end{abstract}


\end{frontmatter}


Understanding nuclear structure and dynamics in terms of the fundamental interactions between protons and neutrons is one of the overarching goals of nuclear science.
To this end, nuclear theory is developing chiral effective field theory (EFT) \cite{Epel09RMP,Mach11PR}, a unified approach to nuclear forces, where two-nucleon (NN), three-nucleon (3N) and higher-body forces are derived within a consistent, systematically improvable framework. 
This approach coupled with parallel advances in ab initio many-body theory \cite{Hage14RPP,incollection,Hebe15ARNPS,Herg16PR,Stro19ARNPS} provides the possibility to link the structure of  nuclei to the underlying symmetries of quantum chromodynamics. 

Neutron-rich oxygen isotopes are particularly fruitful candidates to test  ab initio theory at the interface of the light- and medium-mass regions.
Due to their semi-magic nature, most oxygen isotopes are accessible to many-body approaches amenable to heavier systems, while still being light enough to be treated in quasi-exact methods, such as extensions of the no-core shell model (NCSM). 
First valence-space calculations with NN+3N forces were able to explain, for the first time, the location of the oxygen dripline at $^{24}$O \cite{Otsuka2010}.
More recently, large-space ab initio calculations, where all nucleons are treated as explicit degrees of freedom, have confirmed those early results \cite{Hergert:2013vag,Cipo13Ox,Hebe15ARNPS} and new calculations have even extended dripline predictions to the entire region \cite{Holt19drip}.
Furthermore excitation spectra in oxygen have also been obtained with NN+3N forces, generally yielding agreement with experiment approaching that of state-of-the-art phenomenology \cite{Holt2013,Caes1326O,Bogner2014,Jansen2014}. 
An important next step is to benchmark ab initio theory against other observables which are sensitive to physics beyond what is relevant for excitation energies alone. 
For instance the long-standing problem of quenching of beta decays across the nuclear chart has recently been explained \cite{Gysb19GT}, but electromagnetic properties have only been intermittently studied \cite{Parz17Trans,Ruiz15Ca,Hend18E2,Klos19Ca}. 
In particular, limited data exists for transition rates in the neutron-rich oxygen isotopes. 
In $^{21}$O no experimental information is available on transition strengths, while $\gamma$ decays from bound excited states beyond the first have been reported in \cite{Stanoiu2004} with limited statistics. 

In this Letter, we report first electromagnetic transition rates from low-lying excited states of  $^{21}$O.
We compare our results to predictions from phenomenological shell model and two ab initio many-body methods, the in-medium (IM-) NCSM and the valence-space in-medium similarity renormalization group (VS-IMSRG). Using a number of chiral EFT NN+3N forces, we study the sensitivity of electromagnetic transitions to details of nuclear interactions. 
It should be noted that electromagnetic two-body currents, currently under development, are not included in the ab initio calculations.

The experiment was performed at the National Superconducting Cyclotron Laboratory (NSCL) at 
Michigan State University. A $^{24}$F secondary beam was produced by fragmenting a 
140~MeV/nucleon $^{48}$Ca beam on a 893\,mg/cm$^2$ $^{9}$Be production target. The 
A1900 separator \cite{a1900-1, a1900-2} was used to select and transport the $^{24}$F ions 
(with an energy of 95\,MeV/nucleon, a 2.5\% momentum dispersion and 95\% purity)
to the experimental vault where they underwent reactions on a secondary 2\,mm  
$^{9}$Be target, located
at the target position of the S800 spectrograph \cite{s800}. $^{21}$O was produced via the 
$^{9}$Be($^{24}$F,$^{21}$O+$\gamma$)X 
multi-nucleon removal reaction and identified on an event-by-event basis via energy-loss and 
time-of-flight measurements.
Emitted $\gamma$ rays were detected with the Gamma-Ray 
Energy Tracking In-beam Nuclear Array (GRETINA) \cite{Paschalis2013,Weisshaar2017}.

Data were recorded with two different settings: I) the secondary  
$^{9}$Be target was solely used, and II) a 0.92\,mm $^{181}$Ta degrader was 
mounted at distances of 25\,mm (II/25) or 45\,mm (II/45) downstream of the target using the TRIple PLunger for 
EXotic beams \cite{triplex}. The 
velocity of the $^{21}$O fragments was $v/c \approx 0.41$ after the target 
and $v/c \approx 0.36$ after the degrader. 
To maximize the sensitivity and efficiency of GRETINA, the target position was moved 
upstream by 13\,cm for setting I 
and II/25, covering angles between 20$^{\circ}$ to 70$^{\circ}$, 
and by an additional 20\,mm for setting II/45.
The tracking capabilities of GRETINA were used to determine the interaction sequence and thus the detection angle of the $\gamma$ rays used for the Doppler correction.
The trajectory of the outgoing $^{21}$O fragments was measured with S800 and 
their determined angles and momenta were employed in the Doppler correction on an 
event-by-event basis.

\begin{figure}
    \centering
    \includegraphics[width=0.7\textwidth, trim={47 0 21 20},clip]{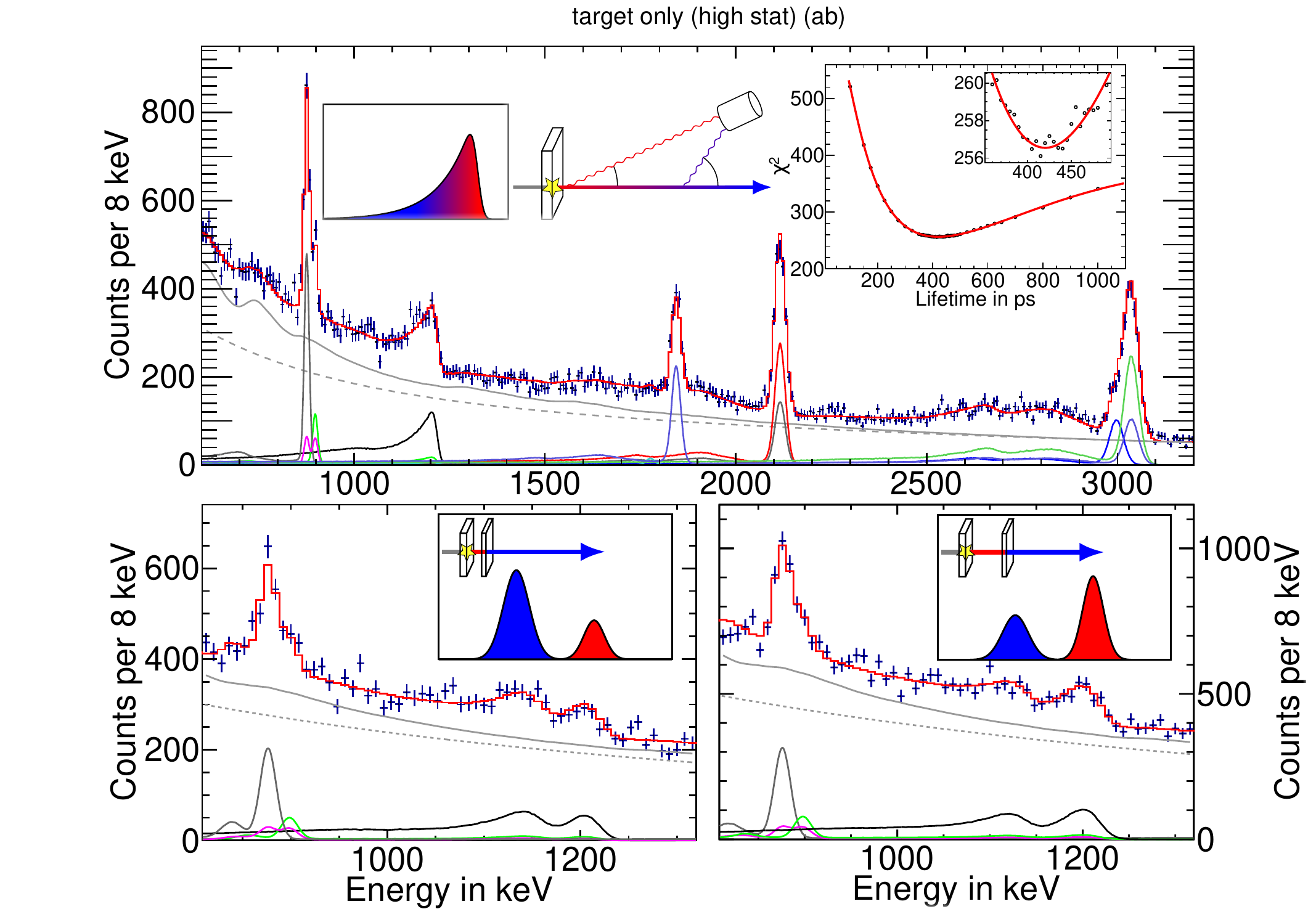}
    \begin{picture}(0,0)
            \put(-155,62){\textsf{\textcolor{blue}{\scriptsize slow}}}
            \put(-140,58){\textsf{\textcolor{red}{\scriptsize fast}}}
            \put(-57,62){\textsf{\textcolor{blue}{\scriptsize slow}}}
            \put(-37,64){\textsf{\textcolor{red}{\scriptsize fast}}}
            \put(-217,160){\textsf{(a)}}
            \put(-200,68){\textsf{(b)}}
            \put(-98,68){\textsf{(c)}}
            \put(-85,160){\textsf{(d)}}
        \end{picture}
        \caption{Comparison of experimental $\gamma$-ray spectra to simulations for $\tau_{1/2^+} = 420$\,ps in settings (a) I, (b) II/25 and (c) II/45. Simulations for different cascades are shown in varying colours. The laboratory-frame background (grey, solid) is shown on top of a double exponential background (grey, dashed). The sum of all simulated spectra (red) has been fitted to the data (blue). The insets illustrate the spectral features of the lifetime measurements. (d) Un-normalized $\chi^2$ distribution for $\tau_{1/2^+}$, combined for all settings, with a feeding lifetime of $\tau_{3/2^+} = 8$\,ps and DRR = 22.6\%, and fitted with a fourth-order polylogarithmic function.
        The inlay zooms around the minimum of the distribution, with a reduced $\chi^2 \approx 1$ (Neyman $\chi^2$ with 265 degrees-of-freedom).}
        \label{spectra}
\end{figure}
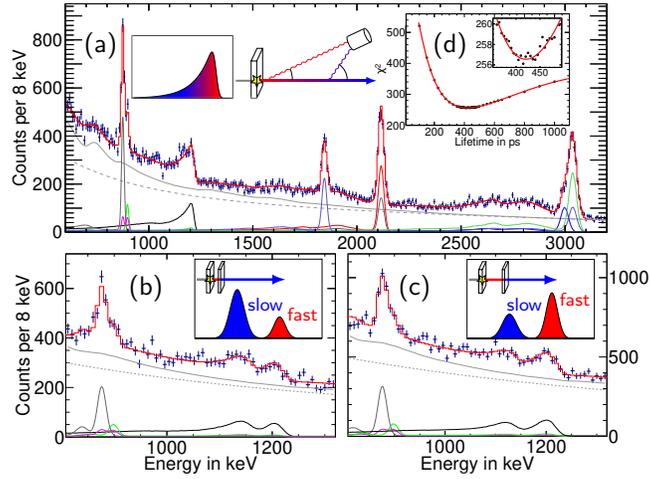

To determine the lifetime of the $^{21}$O $1/2^+$ state, its $\gamma$-ray decay to the $5/2^+$
ground state has been analysed using data from settings I and II. 
In setting I, the lifetime is inferred from the low-energy tail that is generated in the 
$\gamma$-ray spectrum, see Fig.~\ref{spectra}(a). This 
tail results from nuclear levels decaying farther away from 
the target, but being Doppler corrected as if they decay promptly at the target 
position, see, e.g., \cite{Terry2008, Doornenbal2010, Pardo2012, Whitmore2015}.
In setting II, the Recoil Distance Method (RDM) was employed, where $\gamma$ rays emitted before or after the degrader experience different Doppler shifts leading to two laboratory
energies, typically called the fast and slow component of the peak in the $\gamma$-ray spectrum, as can be seen in the insets of Fig.~\ref{spectra}(b) and (c).
The ratio of the number of $\gamma$ rays in the fast and slow peak infers the lifetime
of the state. The RDM for fast beams and its 
implementation at the NSCL is described in \cite{Whitmore2015,rdm-1, rdm-2, rdm-3, 
PhysRevLett.107.102501, PhysRevC.86.044329}.
Since reactions populating the state of interest occur also in the degrader, the degrader-reaction ratio (DRR) has to be taken into account when evaluating the fast and slow components in an RDM measurement.
The DRR has been determined by evaluating $\gamma$-ray transitions with much shorter lifetimes. Since the production mechanism is a multi-nucleon removal reaction, the DRR is assumed to be similar for all transitions.

\begin{figure}[tbp]
 \centering
 \includegraphics[angle=0, width=0.65\textwidth, trim=0 0 0 0, clip=true]{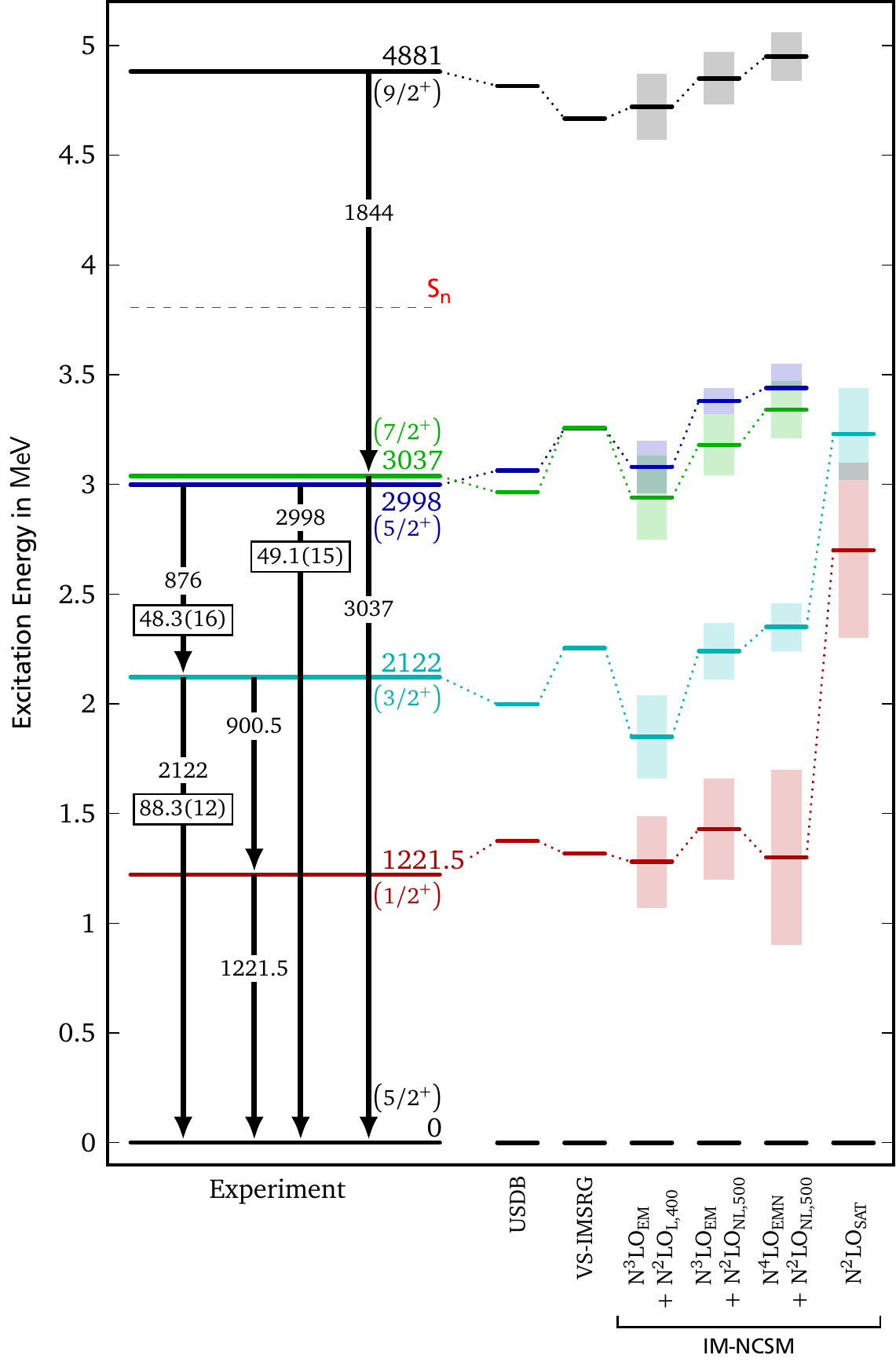}
 \caption{Level scheme of $^{21}$O determined in this experiment and how this compares to
IM-NCSM with four different chiral NN+3N
interactions (see text for details), VS-IMSRG, and phenomenological
shell model with $\rm e_n = 0.45$. Experimental branching ratio (in \%) of $\gamma$-ray decays are marked in rectangles.
The 3037 and 1844\,keV $\gamma$-ray transitions are in coincidence, as well as the 876 and 2122\,keV, placing them firmly on the level scheme. The 900.5\,keV transition has been observed for the first time, allowing the determination of the $\gamma$-ray decay branching ratio of the 2122\,keV level. Ordering of the ($5/2^+$) and ($7/2^+$) states at 3\,MeV has been based on $\gamma$-ray branching-ratio consideration. The colored bands represent theoretical uncertainties (see text for details).}
 \label{theory}
\end{figure}

To determine the lifetime, branching ratios and DRR, the measurements are compared to simulations obtained with Geant4 \cite{Agostinelli2003}, simulating all relevant properties, i.e., detector geometry and response, $\gamma$-ray cascades, lifetimes, beam profile \cite{rdm-1}. 
The energy of the $1/2^+$ state is taken as $1221.5 \pm 2.2$\,keV, the average from two $\beta$-decay experiments \cite{Sumithrarachchi2010, Li2009}.
The energies of the remaining levels are determined from the data.
The experimentally obtained level scheme for $^{21}$O is shown in Fig.~\ref{theory}.
All cascades are simulated separately, 
as well as the neutron induced background from $\rm Ge(n,n')$ and $\rm Al(n,X)$ reactions in GRETINA and the beam-line, respectively.
The neutron induced background is simulated as a combination of non-moving $\gamma$-ray sources at the target position with the respective energies, where the intensities are obtained from the data.
This background is then Doppler corrected and added to the experimental background.
The obtained spectra are summed up with separate normalization factors for each simulation.
Finally the summed spectrum is modified by a double-exponential background and fitted to the measured spectra using the least-squares method.
The free parameters are hereby given by the exponential background, the normalization factor for each setting, and the relative strengths of each of the simulated cascades.
The minimization is performed with all parameters free over the full energy range for all settings simultaneously.
Such simulations are then performed for various DRRs and a $\chi^2$ minimization is performed determining DRR = $22.6\% \pm 1.8$\%.
Using the data from settings I and II and the determined DRR, the lifetime of the $1/2^+$ state is extracted as $\tau_{1/2^+} = 420^{+35}_{-32}\text{(stat)}^{+34}_{-12}\text{(sys)}\,\text{ps}$ in an equivalent minimization process as described for the DRR, resulting in the $\chi^2$ distribution shown in Fig.~\ref{spectra}(d).
The $\chi^2$ is calculated between $1050$\,keV and $1260$\,keV, covering both the full peak and tail for setting I, as well as both fast and slow peaks for settings II.
The statistical uncertainty is given by the un-normalized $\chi^2_{min}+1$ range, while
the systematic uncertainty is dominated by the uncertainty of the DRR and the energy of the first excited state.
To construct a consistent level scheme from the measured $\gamma$-ray energies, the lifetime of the $3/2^+$ state is determined via the centroid-shift method, see, e.g., \cite{Pardo2012, Ralet2017}, to $\tau_{3/2^+} = 8^{+21}_{-8}$\,ps.

\begin{figure}
    \centering
       \includegraphics[height=0.4\textheight, trim={70 230 110 160}, clip]{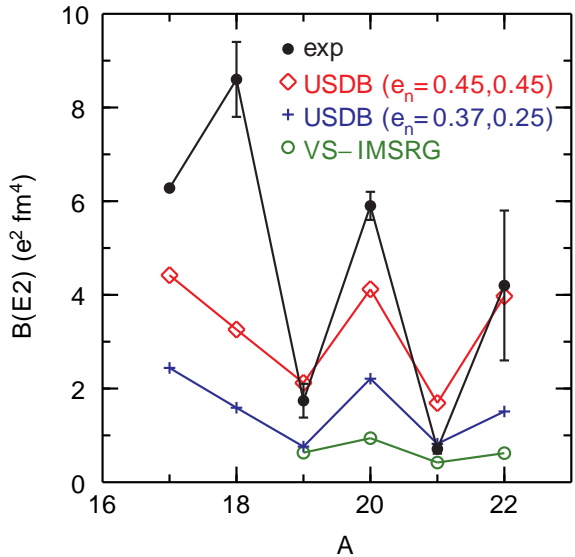}
       \caption{Experimental B(E2) for neutron-rich oxygen isotopes \cite{TILLEY19951,PRITYCHENKO20161} and how this compares with the shell model and VS-IMSRG results. For even and odd isotopes the $2^+ \rightarrow 0^+$ and $1/2^+ \rightarrow 5/2^+$ transition is considered, respectively.}
       \label{babrown}
\end{figure}

We compare the experimental results to phenomenological shell model, and two ab initio methods
using chiral EFT NN+3N interactions, the VS-IMSRG and
the IM-NCSM, see Figs.~\ref{theory},~\ref{babrown} and Table~\ref{results-1}.
 
Within the phenomenological shell model, 
the electric quadrupole B(E2) are obtained with the USDB $sd$-shell Hamiltonian including the neutron $1d_{5/2}$, $2s_{1/2}$, and $1d_{3/2}$ single-particle orbitals \cite{Brow06USD}.
The E2 matrix elements are calculated with harmonic-oscillator radial wavefunctions, and with an effective neutron charge $\rm e_n=0.45$, obtained from a fit to experimental B(E2) values of $sd$-shell nuclei ($\rm A=17-38$) \cite{b}.
In addition, the results for $\rm A  = 17-22 $ are compared to experiment in Fig.~\ref{babrown}.
For nuclei close to $^{16}$O, and in particular for $^{18}$O, the experimental B(E2) are larger than those calculated,
due to the mixing with low-lying states coming from the excitation of protons from the $  p  $ shell to the $  sd  $ shell \cite{c}. 
These core-excited states move to a higher excitation energy for heavier nuclei, as their mixing with the $  sd  $-shell states becomes smaller.
Thus $^{21}$O, free from such state mixing, is ideal to examine the shell-model effective charge.
The overestimated B(E2) in $^{21}$O compared to experiment (see Table~\ref{results-1}) shows
that the neutron effective charge is smaller than
the average value for the $sd$ shell ($\rm e_n=0.45$).
The reason for this was discussed in \cite{c} in terms of microscopic theories for the effective charge,
the latter being orbital and mass dependent. 
The average for the effective
charges for
$  1d-1d  $ and $  1d-2s  $
for $  \rm A=17  $ ($  \rm Z=8  $) and $  \rm A=40  $ ($  \rm Z=20  $) is $\rm e_n=0.50$ \cite{c}, which is close to the
average empirical value of $\rm e_n=0.45$ from \cite{b}
used for the calculations. For $^{21}$O the $1/2^{ + }$ to
$5/2^{ + }$ transition is dominated by the $  1d-2s  $
one-body transition density. The effective charges
for $\rm Z=8$ are $\rm e_n=0.374$ for $  1d-1d  $ and $\rm e_n=0.248$ for $  1d-2s  $.
The reason for the smaller $1d-2s$ effective charge
is that in this case the valence transition density has a node
near the maximum value of the core-polarization
density $k(r)$ in Eq.~(8) of \cite{c}. This results in some
cancellation in the integral whose integrand
is a product of these two densities.
If these values are
used for calculations, the transition strength for $^{21}$O is reduced
to $\rm B(E2)=0.82~e^2fm^4$ and agrees with experiment (see Fig.~\ref{babrown}).
It is interesting to confirm whether these reduced effective charges can reproduce the B(E2) for $^{22}$O. The latter has been measured in \cite{THIROLF200016} with large uncertainties. New experiments aiming at constraining this value, e.g., \cite{Petri-ANL}, will thus add to our understanding of the oxygen isotopes.

The VS-IMSRG~\cite{Herg16PR,Stro19ARNPS,Tsuk12SM,Bogn14SM,Stro17ENO,Stro16TNO} provides a framework to produce ab initio valence-space Hamiltonians, based on NN+3N forces derived from chiral EFT. 
Working in a Hartree-Fock basis, we use the Magnus formulation of the IMSRG \cite{Herg16PR,Stro19ARNPS,Morr15Magnus}, to first decouple the $^{16}$O core energy. 
Then we decouple an $sd$-shell Hamiltonian, using the ensemble normal ordering procedure described in \cite{Stro17ENO}, to include effects of 3N forces between valence nucleons, specifically the five valence neutrons for the $^{21}$O energies and transition rates.
Finally, we use the approximate unitary transformation from the Magnus framework to additionally decouple an M1 or E2 valence-space operator consistent with the valence-space Hamiltonian~\cite{Parz17Trans}. 
In this framework, effective charges are thus not needed, but the effective operator is calculated consistently. 
Unless otherwise specified, all other technical details are the same as in \cite{Parz17Trans,Stro17ENO}. 
The particular input NN+3N interaction used here, EM 1.8/2.0 developed in \cite{Hebe11fits,Simo17SatFinNuc}, begins from a chiral NN interaction at next-to-next-to-next-to-leading order (N$^3$LO) \cite{Ente03EMN3LO} and 3N forces at N$^2$LO. 
This Hamiltonian, fit to few-body data, has been shown to reproduce ground- and excited-state energies across the nuclear chart from the $p$ shell to the tin region \cite{Hebe15ARNPS,Holt19drip,Simo17SatFinNuc,Morr18Sn,Taniuchi19}. 
Indeed, very good agreement between the experimental and VS-IMSRG excited-state energies is observed in Fig.~\ref{theory}.
While E2 transition rates in the $sd$-shell are generally systematically below experiment, owing to the difficulty in capturing the highly collective physics of this transition, the trends typically agree well with experiment~\cite{Hend18E2}. This can also be seen qualitatively in Fig.~\ref{babrown}, where the staggering of E2 strength resembles experiment. For the odd mass cases, in particular $^{21}$O, the agreement with experiment is rather good, while for the more collective
transitions in even-mass isotopes the VS-IMSRG largely underestimates
experiment.

We perform ab initio IM-NCSM calculations in the framework introduced in \cite{Gebrerufael:2016xih}.  
This novel method is a combination of the NCSM with a multi-reference IMSRG evolution of the many-body Hamiltonian that decouples a multi-determinantal reference state, typically an NCSM eigenstate from a small $N_{\max}^{\text{ref}}=2$ reference space, from all Slater determinants outside of this reference space. 
The resulting Hamiltonian is employed in a final NCSM calculation to extract ground and excited states and all relevant observables. The decoupling leads to an extremely fast model-space convergence of the energies. 
For the application to electromagnetic observables in $^{21}$O, two important developments beyond the basic IM-NCSM discussed in \cite{Gebrerufael:2016xih} were necessary: the consistent multi-reference in-medium evolution of the electromagnetic operators, as well as an extension to odd particle numbers via a particle-attachment or particle-removal scheme. 
The details of these extensions are presented in \cite{Vobig2019}. 
We perform IM-NCSM calculations with four different chiral NN+3N interactions to asses the sensitivity of our results to the input Hamiltonian: (i) the N$^2$LO$_\text{SAT}$ interaction \cite{Ekst15sat} ($\Lambda=450\,\text{MeV}$, $\alpha=0.08\,\text{fm}^4$); (ii) the N$^3$LO$_{\text{EM}}+$N$^2$LO$_\text{L,400}$ interaction using the NN force of \cite{Ente03EMN3LO} in combination with a local 3N interaction at N$^2$LO with reduced cutoff \cite{Roth:2011vt} ($\Lambda=400\,\text{MeV}$, $c_D=-0.2$, $\alpha=0.08\,\text{fm}^4$); (iii) the N$^3$LO$_{\text{EM}}+$N$^2$LO$_\text{NL,500}$ with the same NN force but an updated 3N interaction with a nonlocal regulator ($\Lambda=500\,\text{MeV}$, $c_D=0.8$, $\alpha=0.12\,\text{fm}^4$); and (iv) the N$^4$LO$_{\text{EMN}}+$N$^2$LO$_\text{NL,500}$ with a recent NN interaction at N$^4$LO \cite{Entem:2017gor} plus a 3N interaction at N$^2$LO with nonlocal regulator ($\Lambda=500\,\text{MeV}$, $c_D=-1.8$, $\alpha=0.16\,\text{fm}^4$). 
We use the particle-removed calculation at the largest available $N_{\max}=6$ as nominal result and the difference to the particle-attached calculations and the residual $N_{\max}$-dependence to quantify the uncertainty of the many-body calculation. 
Fig.~\ref{theory} shows that the IM-NCSM calculations mostly provide a consistent description of the low-lying spectrum in very good agreement with experiment, 
except for the N$^2$LO$_\text{SAT}$ interaction. 
While the latter includes information beyond the few-body sector, particularly oxygen energies and radii, into the fit, it nonetheless produces a $1/2^+$ state over 1~MeV higher than experiment. 
The B(E2) transition strength from the first excited $1/2^+$ to the ground state, shown in Table~\ref{results-1}, indicate interesting differences, even among the interactions that provide a consistent excitation spectrum. 
The N$^3$LO$_{\text{EM}}+$N$^2$LO$_\text{L,400}$ interaction using a local 3N regulator gives a significantly small B(E2) compared to the other interactions that use nonlocal regulators.
This shows that the E2 observables measured here provide a good test for chiral interactions that goes beyond the aspects probed by the excitation energies alone.
A systematic study of the oxygen isotopes in the IM-NCSM is under way \cite{Roth2019}.

\begin{landscape}
\begin{table}
\newcommand\jp[1]{\textsf{\ensuremath{\tfrac{#1}{2}^{+}}}}
\newcommand\gs{\textsf{\ensuremath{\tfrac{5}{2}^{+}\hspace{-4pt}\text{\footnotesize gs}}}}
    \caption{Comparison of experiment and theory. The experimental B(E2) is deduced from the lifetime ($\tau$) of the state. The theoretical branching ratio (BR) is derived using the theoretical B(E2) and B(M1) and the experimental transition energies. The theoretical $\tau_{1/2^+}$ is calculated using the experimental transition energy.}
    \label{results-1}
    \centering
    \begin{tabular}{l *8{c}}\hline\hline
        & \multicolumn{1}{c}{$\tau$ [ps]}
        & \multicolumn{1}{c}{BR [\%]} 
        & \multicolumn{3}{c}{B(E2) [e$^2$fm$^4$]} & \multicolumn{2}{c}{B(M1) [$10^{-3} \mu_\text{N}^2$]} \\
            \cline{2-2} \cline{3-3} \cline{4-6} \cline{7-8}
        &  $\jp{1}$
        & $\jp{3}{\rightarrow}\ \jp{1}$
        & $\jp{1}{\rightarrow}\ \gs$ & $\jp{3}{\rightarrow}\ \jp{1}$ & $\jp{3}{\rightarrow}\ \gs$
        & $\jp{3}{\rightarrow}\ \jp{1}$ & $\jp{3}{\rightarrow}\ \gs$ \\
        \cline{2-2} \cline{3-3} \cline{4-6} \cline{7-8}
        Experiment
                & $420^{+35\ +34}_{-32\ -12}$    
                & 11.7$\pm$1.2  
                & $0.71^{+0.07\ +0.02}_{-0.06\ -0.06}$ & &  \vspace{1mm}\\
        USDB 	   
                & 176   
                & 20.3  
                & 1.69 & 2.06 & 3.54  & 5.6 & 0.6 \\
        VS-IMSRG      
                & 704 
                & 12.7 
                & 0.42 & 0.61 & 0.55 & 5.3 & 2.6  \\
        N$^2$LO$_\text{SAT}$
                &  $444^{+128}_{-81}$  
                &  $4.6^{+5.5}_{-2.6}$ 
                & 0.67$\pm$0.15 & 0.63$\pm$0.05 & 0.70$\pm$0.06 & 3.0$\pm$1.2 & 4.6$\pm$1.9\\  
        N$^3$LO$_\text{EM}$+N$^2$LO$_\text{L,400}$
                    & 804$\pm$22 
                    & $12.5^{+3.2}_{-2.5}$	
                & 0.37$\pm$0.01 & 0.47$\pm$0.05 & 0.55$\pm$0.02 & 9.8$\pm$1.0 & 5.1$\pm$0.8  \\
        N$^3$LO$_\text{EM}$+N$^2$LO$_\text{NL,500}$
                      &	 $488\pm25$  
                      &  14.1$^{+2.4}_{-1.9}$
                & 0.61$\pm$0.03 & 0.74$\pm$0.08 & 0.77$\pm$0.04 & 8.4$\pm$0.4 & 3.7$\pm$0.5 \\
        N$^4$LO$_\text{EMN}$+N$^2$LO$_\text{NL,500}$
                     & $513^{+94}_{-69}$ 
                     &  15.5$\pm$1.3 
                & 0.58$\pm$0.09 & 0.70$\pm$0.07 & 0.77$\pm$0.06 & 8.2$\pm$0.5 & 3.2$\pm$0.1\\\hline\hline
    \end{tabular} 
\end{table}
\end{landscape}

In summary, the low-lying structure of $^{21}$O was studied at the NSCL using GRETINA coupled to the S800 
spectrometer. The lifetime of the first (and second) excited state was measured 
for the first time, as well as $\gamma$-ray branching ratios from higher-lying states. Our experimental 
results are compared to ab initio VS-IMSRG and IM-NCSM predictions,
demonstrating that E2 observables provide an interesting testing ground for 
chiral interactions and many-body methods that goes beyond the aspects probed by the excitation energies alone. 
Indeed, comparison of our experimental results with IM-NCSM calculations using different chiral NN+3N interactions suggests that interactions derived with nonlocal 3N regulators better capture the electromagnetic transition rates.
Comparison of experiment with phenomenological shell model demonstrates that neutron-rich oxygen isotopes, and $^{21}$O in particular, are prime candidates to study shell-model effective charges, 
since their low-lying structure is free from core excitations.
\section*{Acknowledgments}
We acknowledge A.~O.~Macchiavelli and S.~R.~Stroberg for enlightening discussions.
This work was supported by the Deutsche Forschungsgemeinschaft (DFG, German
Research Foundation) -- Projektnummer 279384907 -- SFB 1245,
the Royal Society under contract number UF150476, 
the UK STFC under contract numbers ST/L005727/1 and ST/P003885/1, 
by the National Science Foundation (NSF) under Grants No. PHY-1102511, No. PHY-1565546, and No. PHY-1811855,
by the Department of Energy (DOE) National Nuclear Security Administration
and through the Nuclear Science and Security Consortium under Awards No. DE-NA0000979.
GRETINA was funded by the DOE, Office of Science. Operation of the array at NSCL was supported by DOE under Grant No. DE-SC0014537 (NSCL) and DE-AC02-05CH11231 (LBNL).
TRIUMF receives funding via a contribution through the National Research Council Canada. This work was also supported in part by NSERC.
Numerical calculations have been performed at the Lichtenberg HPC cluster at the TU Darmstadt and within allocation of computing resources from Compute Canada and on the Oak Cluster at TRIUMF managed by the University of British Columbia department of Advanced Research Computing (ARC). 


\bibliography{o21}

\end{document}